# Supramolecular physics of ambient water

Alexander Kholmanskiy

Science Center «Bemcom», Moscow, Russia

allexhol@ya.ru, http://orcid.org/0000-0001-8738-0189

**Abstract**

In temperature range from 0 °C to ~100 °C, abnormality of ambient water properties, at normal pressure, are mainly defined by the physic of hydrogen bonds in supramolecular structures (SMS). Application of Arrhenius approximations and modification of temperature dependences (TDs) for 15 physical characteristics of water made it possible to define their activation energies and to differentiate the contributions of equilibrium thermal processes and those of SMS reconfigurations. Reactions of hydrogen bonds breakage and those of hexagonal ice-like clusters transformation limit TDs of viscosity and rotation-translational self-diffusion. Limitation of degrees of freedom of these motions by the effect of anisotropic external factors leads to the reduction of activation energy for TDs of compressibility, sound velocity and thermal conductivity nearly sixfold. Equality of absolute values of activation energies, having opposite signs, for the thermal and configurational components of TD, in its point of extremum, is the condition of TDs' extrema for volumetric density, heat capacity at constant pressure, compressibility and sound velocity. In this case, space-time correlation of water dynamics takes place, on the SMS level, followed by constant-energy transition between metastable SMS phases. Fluctuations of O-H vibration frequency play the role of the factor that controls the SMS dynamics.

**Keywords:** water anomalies; Arrhenius approximations; hydrogen bond, ice-like clusters.

## 1. Introduction

Owing to water, the hierarchy of living systems with homo sapiens at the top has appeared and evolved, on the Earth [1-3]. That is why abnormalities of liquid water properties, at normal atmospheric pressure, can be regarded as a display of Anthropic Principle, on the molecular level. It concerns, first of all, the specific features of temperature dependences (TDs) of ambient water properties at temperature (T) in the range of 0 °C to 100 °C. In these conditions, TDs of volumetric density ($\rho$, 4 °C), molar volume (v, 4 °C), specific heat capacity at constant pressure ($C_p$, 35 °C), thermal conductivity coefficient ($\varkappa$), isothermal compressibility ($\gamma$, 46 °C), sound velocity (V, 75 °C), dynamic viscosity ($\eta$), dielectric constant ($\varepsilon$), coefficient of self-diffusion (D), chemical shift ($\delta$), times of spin ($T_1$) and dielectric constant of relaxation ($\tau_D$) behave untypically, and some of them have either extrema, for different T values ($T_E$ are presented in brackets) or kink of curve at 25 °C [4-9]. In spite that extensive experimental and theoretical research of structure and dynamics of liquid water has been made, molecular mechanism of the origin of abnormal behavior of water TDs and their differentiation is not clearly known, so far [8, 11,12].

Morphological dominant of the three-D molecular structure of water is that of electron orbital geometry of oxygen atom, in free $H_2O$ molecule which is close to tetrahedral one [6]. Such geometry tends to energy minimization by forming four tetrahedral hydrogen bonds (HBs). This configuration features the highest possible stability in a structuree of ice-like hexagonal clusters ($W_6$), at T = 0 °C [8, 13]. Absorption band for $W_6$ is in the range of 200 $cm^{-1}$ to 250 $cm^{-1}$ (see Fig. 1) The nature of motions and bonds in liquid water can be reconstructed based on specific variations in its spectral parameters and radial distribution functions $g_{oo}(r)$, $g_{oH}(r)$, $g_{HH}(r)$ (see Fig. 1a). For this purpose, stationary and pulse methods of IR, Raman, X-ray and neutron spectroscopies are applied along with computer simulations of molecular dynamics [10-22]. The existence of two metastable liquid phases, those of high-density liquid (HDL) and low-density liquid (LDL), is usually assumed to explain the abnormal behavior of water characteristics ($\rho$, v, $C_p$, $\gamma$), in normal ambient conditions [10, 11, 12, 13, 14, 15, 16, 17, 18, 19, 20, 21]. It was found out with the help of X-ray spectroscopy and x-ray Raman scattering methods that [11, 22], below 25 °C, structures with four tetrahedral HBs, typical for hexagonal ice (Ih), dominate in the first coordination shell of LDL water [13]. Upon heating from 25 °C to 90 °C, 5% to 10% of these bonds turn into two hydrogen–bonded configurations with one strong donor and one strong acceptor HB which makes LDL water transform into HDL water. The share of 20% to 40% of tetrahedral configuration of LDL water, in normal ambient conditions, was estimated from experiments while that calculated using molecular dynamics simulations was from 60% to 80% [10, 17, 22].

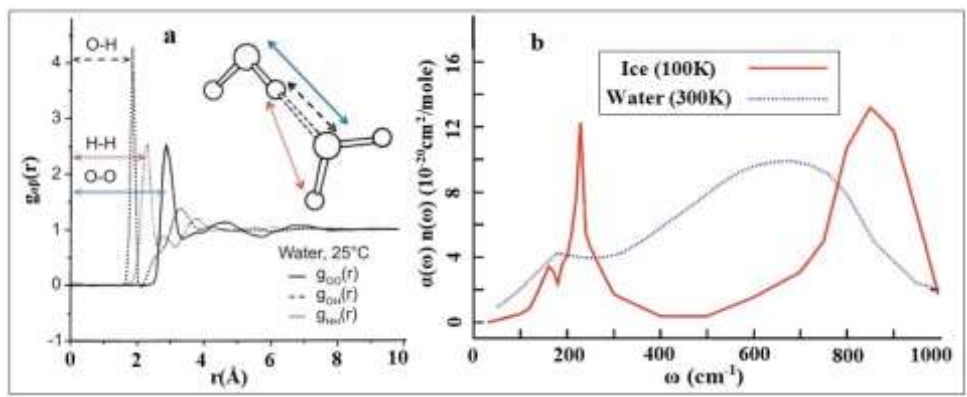

**Fig. 1.** (a) Partial pair distribution functions obtained from neutron diffraction. Figure from [16]. (b) Experimental far-IR absorption spectra of ice and liquid water. Figure from [14].

The assumption was made in works [5, 8, 9, 10, 13, 23, 24, 25, 26, 27, 28] that molecular mechanism of abnormalities in water characteristics is based on the physics of dynamic supramolecular structures (SMS) that are associated with the corresponding 'flash-like' configurations of HBs lattice structure and with correlated molecule motions. The contribution of configurational degrees of freedom, in SMS, to the internal energy of liquid water has the same order of magnitude as that of mechanical oscillations [9]. The density of HBs, their energy and lifetime vary in a rather wide range in SMS. These variations make their effect on the intermolecular interactions and on the dynamics of SMS based on the constrained rotation-vibrational and translational molecular motions. Their energy range is bounded from above by HB energy value ($E_H$) [5]. In thermally static condition when T=constant, water is a kind of storage for energy quanta whose effect appears in Brownian motion. Their energy spectrum corresponds to that of blackbody radiation. The wavelength in the maximum of this band is defined by Wien displacement law [24]:

$$\lambda_{max} (m) = 0.0029/T.$$

This wavelength band for T>240 K will overlap with IR absorption spectrum of water in the range of wave number ($\omega$) below 1000 cm$^{-1}$ (see Fig. 1).

In the range of 273 K<T<373 K, liquid water structure consists of free and 'capsulated' molecules, hexagonal clusters (mainly those in form of $W_6$), various cavities, linear chains and three-D HB lattices (see Fig. 2). Extremal points $T_E$ of TD characteristics of water can be considered as critical ones [30] in which phase transitions between various SMS configurations take place. To describe the physics of these transitions, the formalism of non-equilibrium thermodynamics of phase transitions for clusters is applicable in which configurational excitation can be expressed with the use of transformation and diffusion of cavities in cluster structure with activation energy ($E_A$) [31].

Water inherits its cavitated structure of hexagonal ice Ih while melting, and $W_6$ clusters dominate in the structure of cavities, in temperature range up to 25 $^o$C [5, 6, 8, 9, 22, 24, 32]. Water structure parameters and its rotation-translational spectra depend on T. The order of magnitude of $E_A$ for SMS restructuring can be deduced from Arrhenius approximations (F-approximations) [5, 8], for these TDs and, thereof, to discover a possible nature of molecular motions. In a metastable state, at T<273K, self-diffusion of supercooled water falls considerably, and conformational degrees of freedom get 'frozen', while the role of impurities and gases grows, in terms of water crystallization initiation [20]. In this case, the mathematical certainty of F-approximations for TD characteristics of water and the interpretation adequacy of the molecular motions' nature decrease, as well as that of clusters' structure [33, 34, 35]. At the same time, noises decrease in supercooled water state, and thus the accuracy of spectral methods grows [20]. That is why extrapolations of F-approximations for TD characteristics of water into the range of T<273 may yield useful information concerning the physics of water, on a qualitative level.

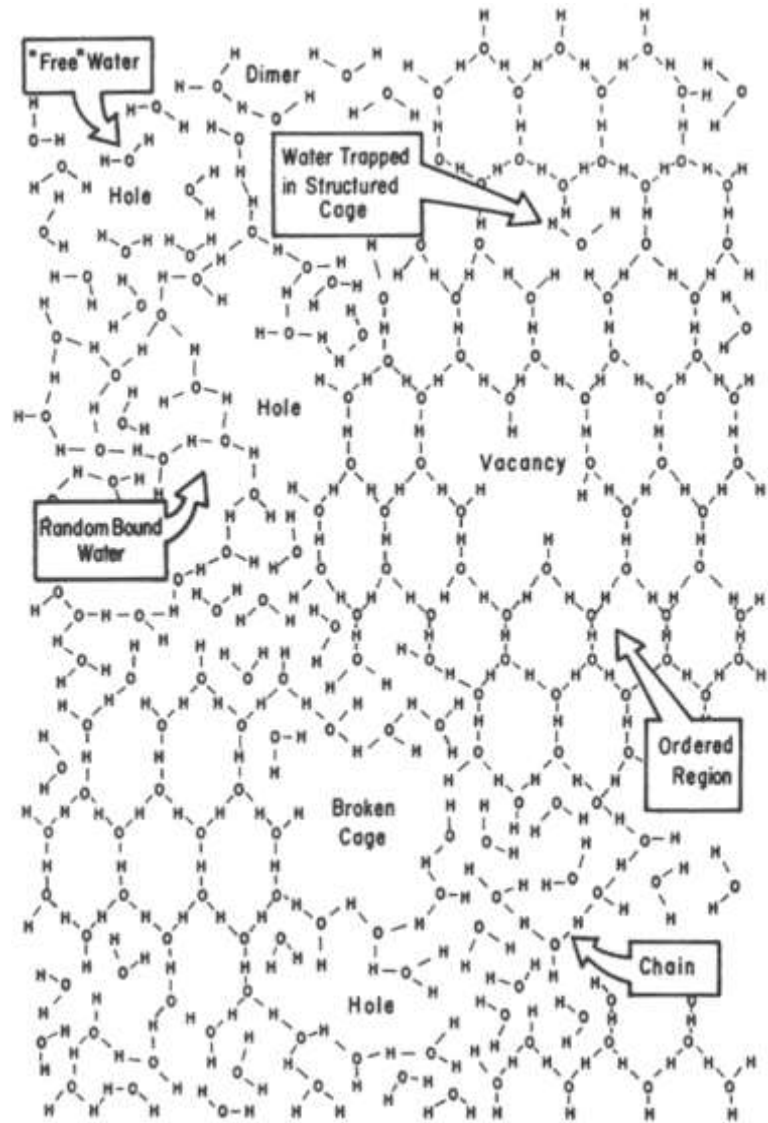

**Fig. 2.** The liquid water-structure complexity is reproduced in a two-dimensional plot for the sake of visualizing some of its characteristics [29].

In this work, Arrhenius approximations for temperature dependences of water's physical characteristics and its structure parameters have been studied in order to define the physical nature of molecular motions and intermolecular bonds responsible for the abnormalities of liquid water properties, at ambient conditions.

**2. Materials and methods**

Empirical data for TDs of water characteristics were imported from published sources. References to these materials are given in figure captions and in tables. Graphs were digitalized with the use of 'Paint' computer application, when necessary. The borders of the temperature range for TDs were defined based on those of available data varying in the T limits of –30 °C to 100 °C. The entire range of T was divided in T-intervals, for certain characteristic of water, where corresponding $T_E$ values were the limiting points. 'MS Excel' application was used to plot TDs and their $F_A$-approximations. The extent of proximity of value $R^2$ to 1 was chosen as the reliability criterion, for F-approximations, in various T-intervals. To transform the exponent of F-approximation into J/mol units their numerical values were multiplied by gas contact R (8.3, J·mol$^{-1}$·K$^{-1}$). Error in calculations for activation energy was determined by the accuracy of either estimated or experimental TDs of corresponding water characteristics and its structural parameters. Typical deviations of values are presented in the tables.

Dependence of water characteristics on T is normally modeled by exponents whose arguments and coefficients depend on T [36]. In paper [5], the empiric TD was represented in the following form:

$$TD = T^{\pm\beta}\exp(\pm E_R/RT), \qquad (1)$$

where $\beta = 0, 1/2, 1, 2$, and $E_R$ is activation energy in $F_R$-approximation, for modified TD* obtained by dividing TD by power function $T^{\pm\beta}$:

$$TD^* = TD/T^{\pm\beta},$$

$F_A$-approximation of TD in form $\exp(\pm E_A/RT)$ can be calculated from (1) by substituting function $T^{\pm\beta}$ with its $F_T$-approximation $-\exp(\pm E_T/RT)$:

$$F_A = F_T F_R = \exp(\pm E_T/RT)\cdot\exp(\pm E_R/RT) = \exp(\pm E_A/RT).$$

Therefore,

$$\pm E_A = \pm E_R \pm E_T.$$

The value and the sign of β, for each water characteristic were selected with the account of known relationships between them, and the adequacy of selection was proved by the high certainty of $F_R$-approximation for TD* function, in the corresponding T-intervals. For translational and rotational self-diffusion of water, in normal conditions, the following Stokes-Einstein relationship is valid for translational and rotational self-diffusion in water, in normal conditions:

$$D\eta \propto T.$$

For η this relationship can be fulfilled for β=0, in case that β=1 is chosen for D [5]. For TDs of O-H stretching vibration frequency ($\nu_{OH}$), value β=0 was chosen for 3620 cm$^{-1}$ (mode $f_1$) and for 3260 cm$^{-1}$ (mode $f_4$) [26].

Considering that physical bodies expand with growing T, the values $\beta = 1$ and $\beta = -1$ were selected, respectively, for v and water volumetric density, from relationships:

$$Pv \propto T \text{ and } v = \mu\rho^{-1}$$

($\mu$ is molar mass of water). With the account that $\rho \propto T^{-1}$, the values of $\beta$ for $\gamma$ and V were adjusted to the following known [7, 9] expression:

$$V = (\gamma \rho)^{-1/2}.$$

Values $\beta=2$ and $-0.5$, for $\gamma$ and V, respectively, were found to be appropriate ones [5]. Value of $C_P$ depends directly on T. That is why $\beta=1$ was chosen for this parameter [5]. Assuming that propagation of heat is responsible for energy transfer and for elastic vibrations in liquids, the following formula was deduced for thermal conduction coefficient ($\varkappa$) [37]:

$$\varkappa \propto \frac{C_p \rho^{4/3}}{\mu^{1/3}}.$$

By substituting $\beta$ for $C_P$ and $\rho$ we obtained $\beta = -1/3$ for $\varkappa$. Dielectric constant ($\varepsilon$) is connected with refraction index of water (n) by expression $\varepsilon=n^2$ and it depends on $T^{-1}$ [9]. Therefore, values $\beta= -1$ and $\beta= -1/2$ were chosen for $\varepsilon$ and n, respectively. While selecting values of $\beta$ for parameters of water microstructure, the fact was considered that radiuses of coordination spheres on the radial distribution function $g_{oo}(r)$ (see Fig. 1) measure up against water volume. It means that $\beta = 1$, for radiuses $r_{oo}$(Å). Variation in local density of water can be estimated based on the height of peak on the curve for the first coordination sphere of water [9, 16]. That is why value $\beta = -1$ was chosen for $g_{oo}(r)$.

The influence of surplus external pressure ($P^+$) on TDs of certain characteristics of water was studied based on the results of works [7, 17, 38, 39, 40, 41, 42]. The following formula was applied to calculate the distance between ions and molecules in water [43]:

$$L = 11.8\, C^{-1/3} \text{ Å}, \qquad (2)$$

where C is concentration, in mol/l. For pure water, L= 3.1 Å, for C=55.6 mol/l which is close to the radius of the first coordination sphere $r_{oo}$=2.8 Å [9].

### 3. Results

Plots for TDs of water characteristics, as well as those of structural and spectral parameters along with their F-approximations are given in Figures 3 to 17. Extremal points ($T_E$ °C) of parameters $\beta$, $E_A$, $E_T$, $E_R$ for functions $F_A$, $F_T$, $F_R$ for micro-characteristics are presented in Table 1 while those for macro-characteristics are given in Table 2.

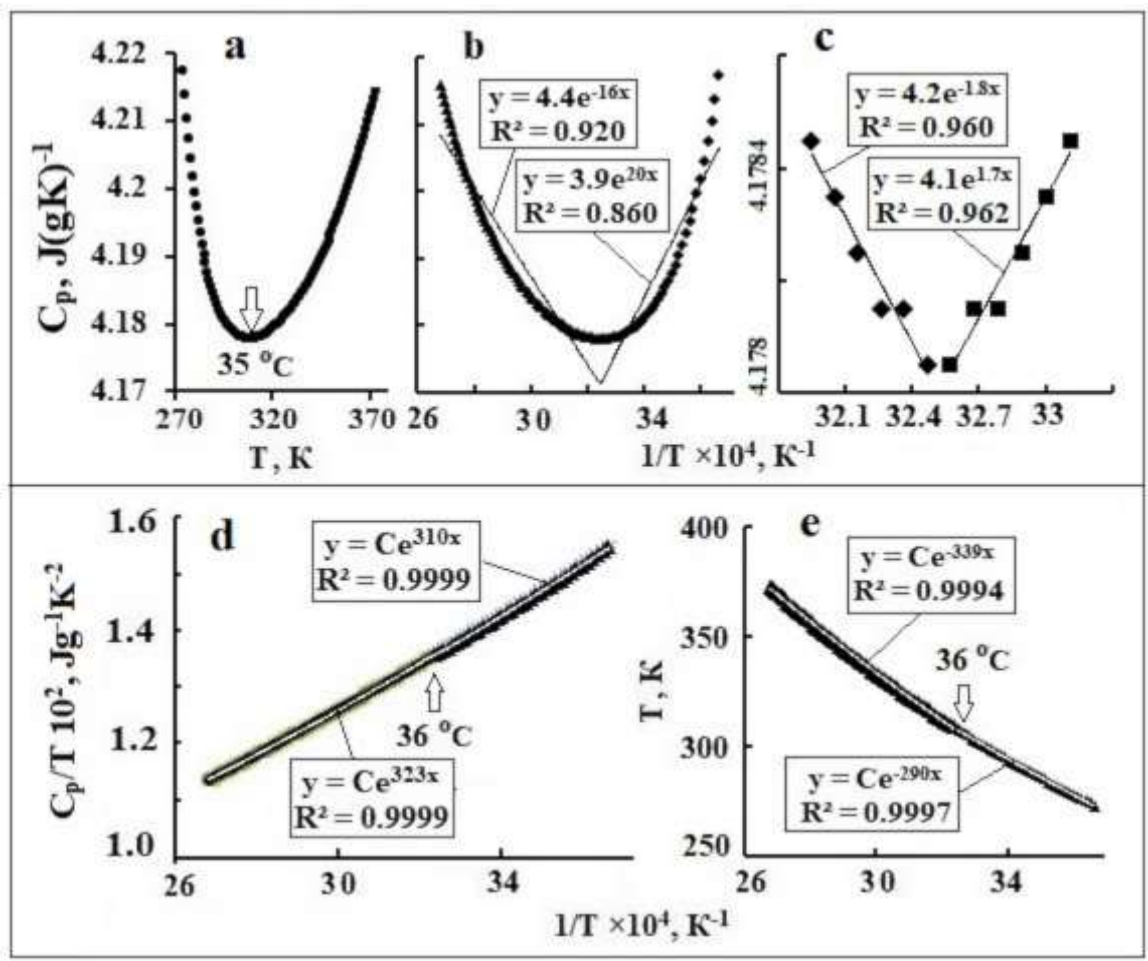

**Fig. 3.** Temperature dependence of water specific heat capacity ($C_p$) on T (a) and on 1/T (b, c); $F_A$-approximations (b, c), $F_R$-approximation (d); dependence T от 1/T и ее $F_T$-approximation (e). Lines of trends are colorless. Data on $C_P$ were imported from [5, 8].

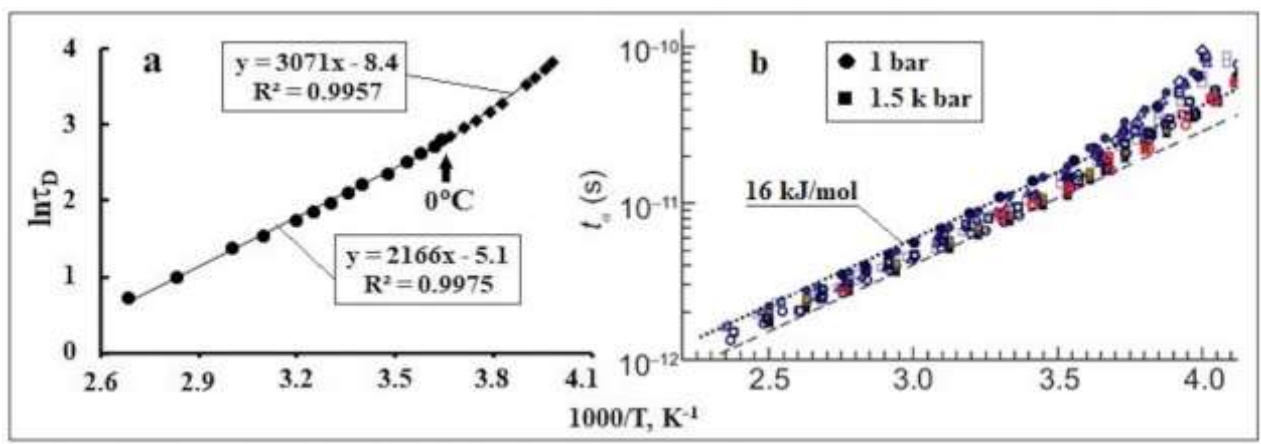

**Fig. 4.** (a) $F_A$-approximation for TD of water dielectric relaxation time ($\tau_D$), initial data imported from [5]. (b) Arrhenius dependence for lifetime of hydrogen bonds ($t_\alpha$) and characteristic times of dynamic processes in liquid water, for various pressures, obtained by NMR method. Adapted figure from [44].

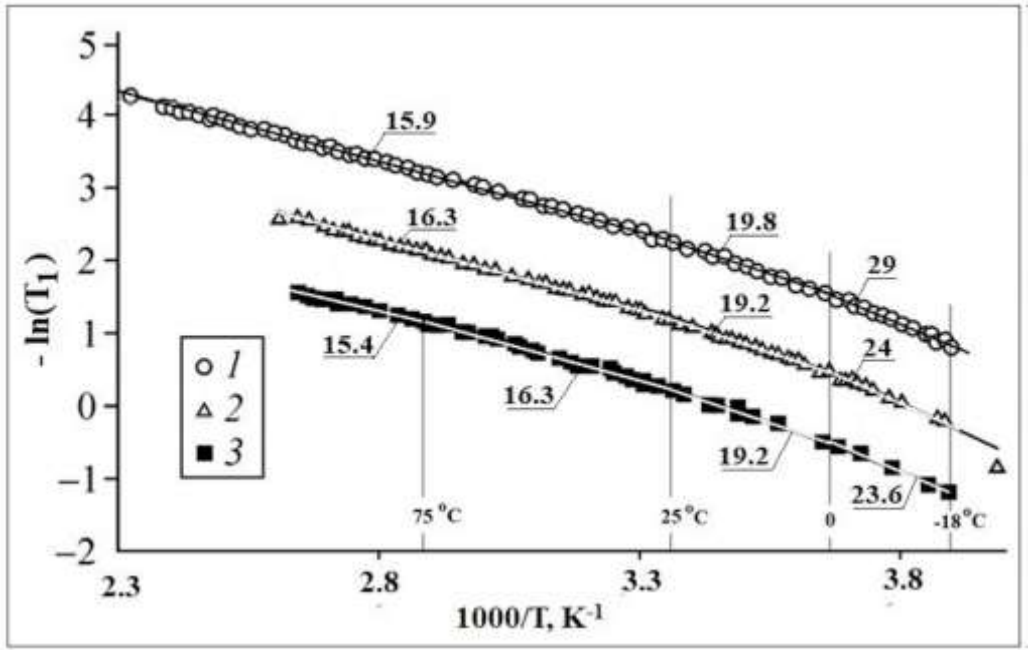

**Fig. 5.** Temperature dependences of the logarithm of spin-lattice relaxation time ($T_1$) and their $F_A$-approximations (lines of trends are colorless); 1 – distilled water, 2 – sea water (~0.6 mol/l NaCl), 3 – NaCl solution (0.5 mol/l). Adapted figure from [36].

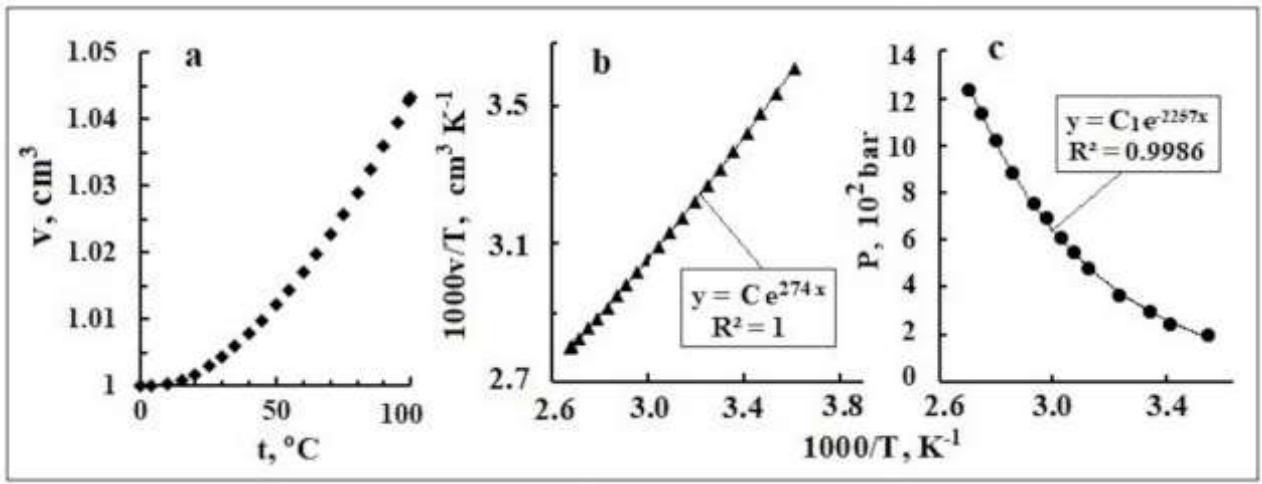

**Fig. 6.** (a) Temperature dependence of water specific (molar) volume (**v**); (b) its $F_R$-approximation. (c) Dependence of pressure (**P**) on $1/T$, for constant specific volume (0.99 cm$^3$g$^{-1}$) and its $F_A$-approximation. Initial data for **v** and **P** from [7].

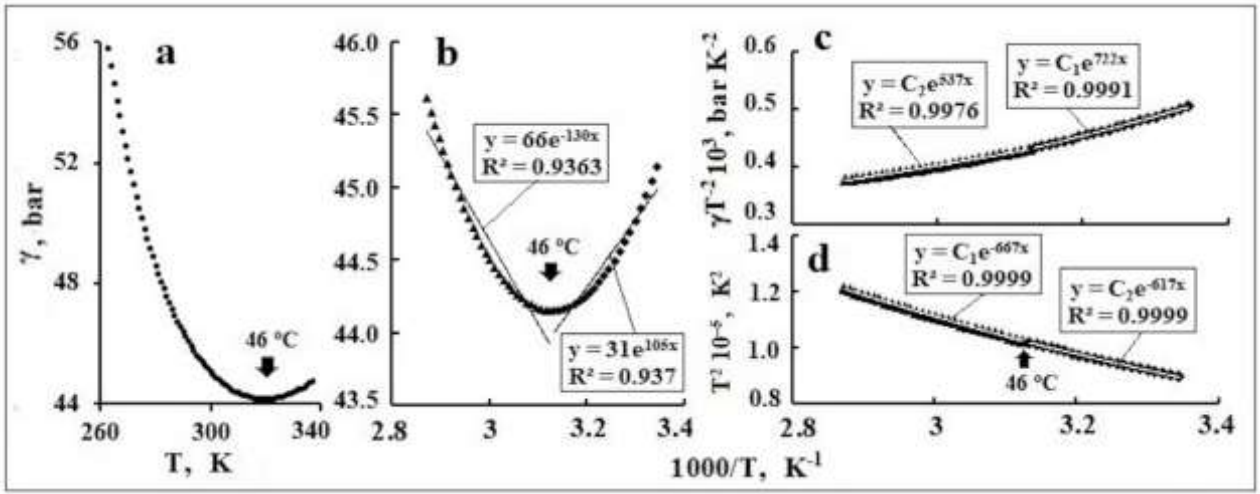

**Fig. 7.** (a) Temperature dependence compressibility (γ); (b) its $F_A$-approximation and (c) $F_R$-approximation; (d) Зависимость $T^2$ от $1/T$ и ее $F_T$-approximation. Data on γ(T) from [5]. Arrows show $T_E$. Lines of trends are colorless.

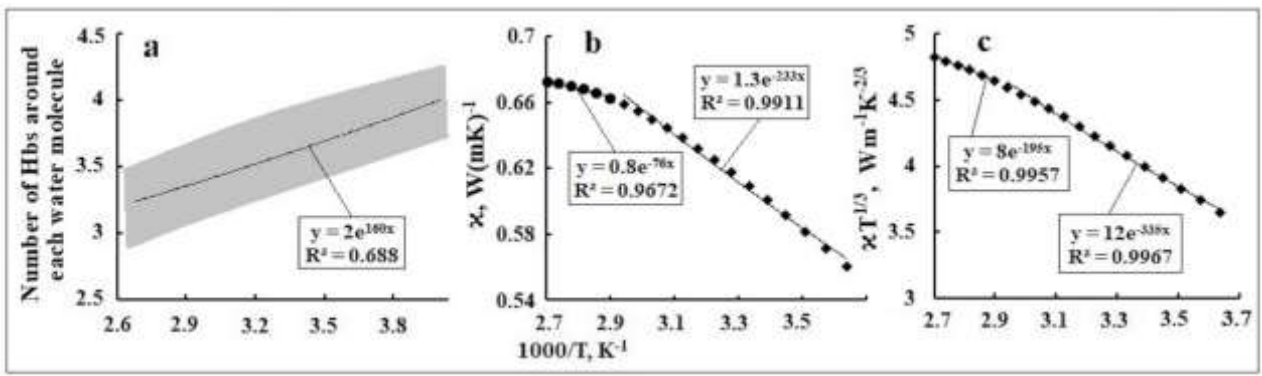

**Fig. 8.** $F_A$-approximation number of hydrogen bonds around each water molecule (a). Adapted figure from [7]. Dependences of the thermal conductivity coefficient (ϰ) (b) and $ϰT^{1/3}$ (c) on $1/T$ and their $F_A$- and $F_R$-approximations. Initial data from [45].

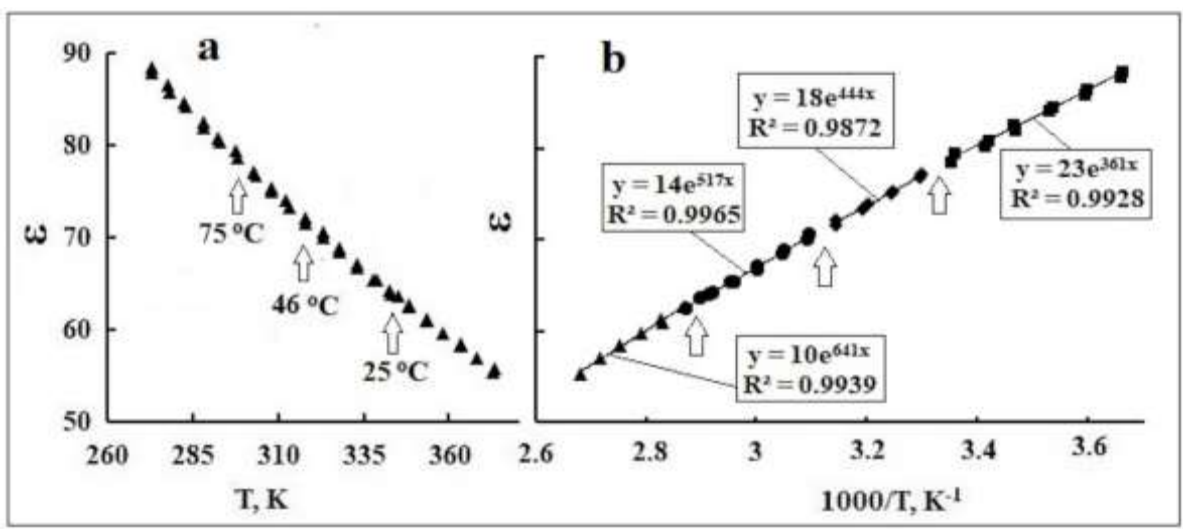

**Fig. 9.** Temperature dependence of dielectric constant (ε) of water (a); its $F_A$- and $F_R$-approximations (b, c). Initial data from [9, 46]).

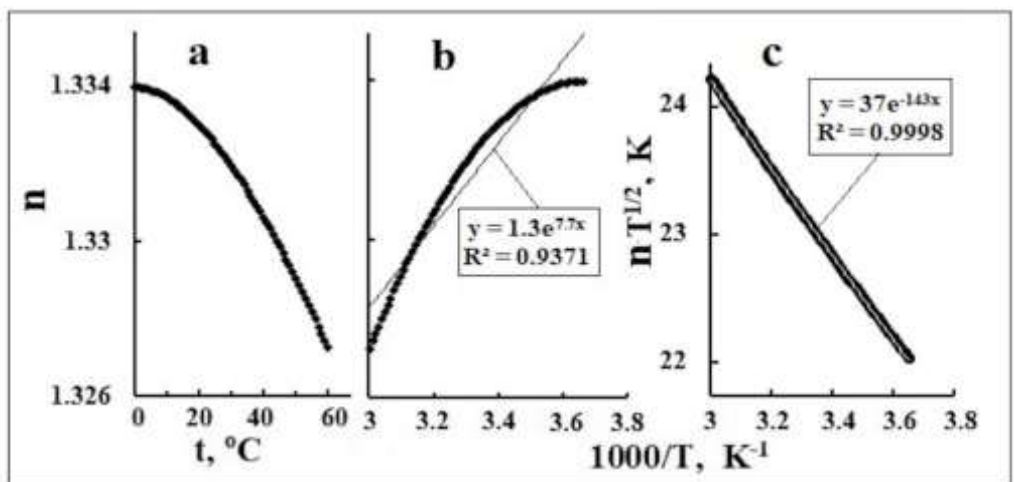

**Fig. 10.** (a) Temperature dependence of the refractive index of water (n) at a wavelength of 589 nm; (b, c) $F_A$- and $F_R$-approximations. Initial data from [47].

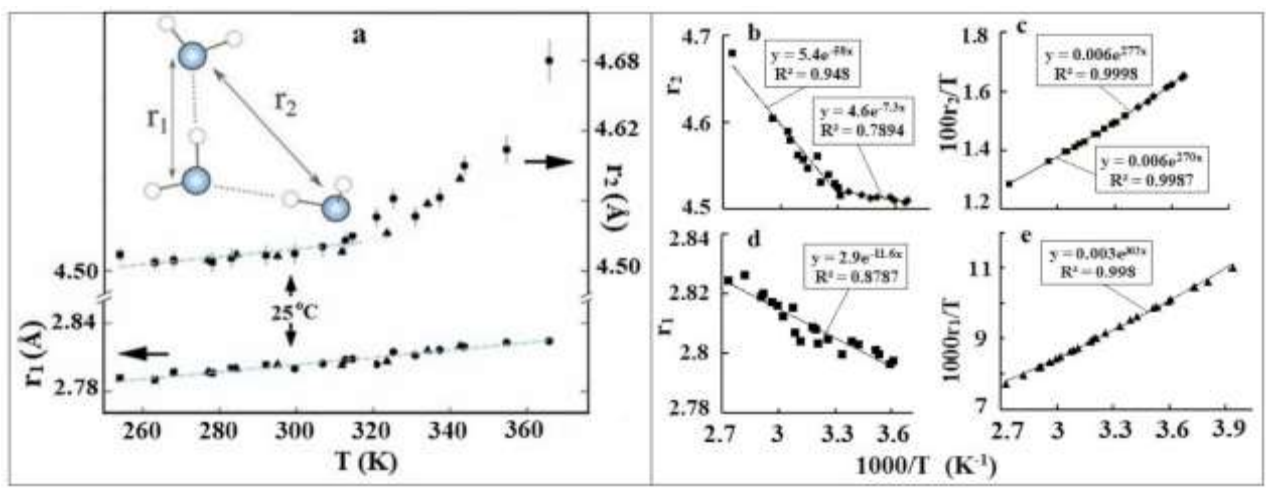

**Fig. 11.** (a) Temperature dependence of the first ($r_1$) and second ($r_2$) peak in $g_{OO}(r)$ function from neutron scattering $D_2O$; (b, d) their $F_A$-approximations; (c, e) their $F_R$-approximations. Adapted figure (a) from [19].

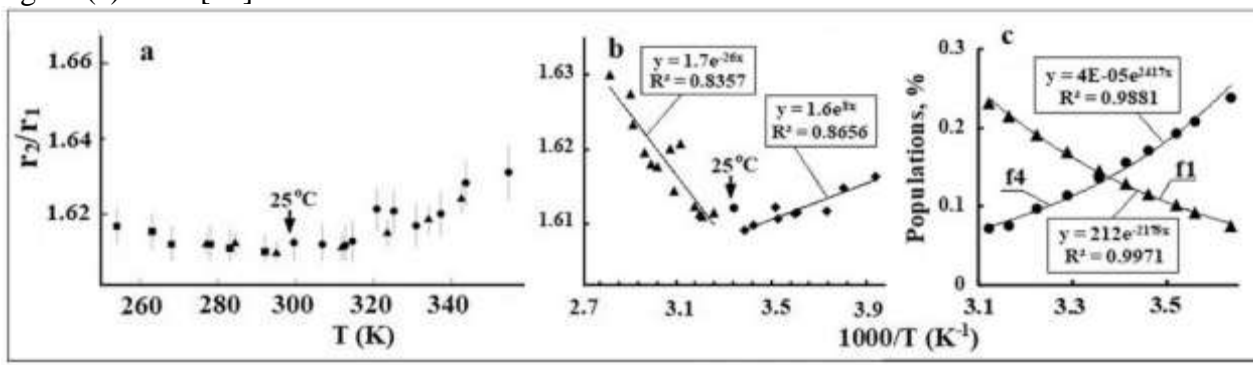

**Fig. 12.** (a) Temperature dependence of the relationship first ($r_1$) and second ($r_2$) peak in $g_{OO}(r)$ function for $D_2O$ ($r_2/r_1$); (b) their $F_A$-approximations. Adapted figure (a) and initial data from [19]. (c) Temperature evolutions populations of O-H stretching modes $f_1$ (the shoulder 3620 cm$^{-1}$) and $f_4$ (the shoulder 3260 cm$^{-1}$) and their $F_A$-approximations. Initial data from [26].

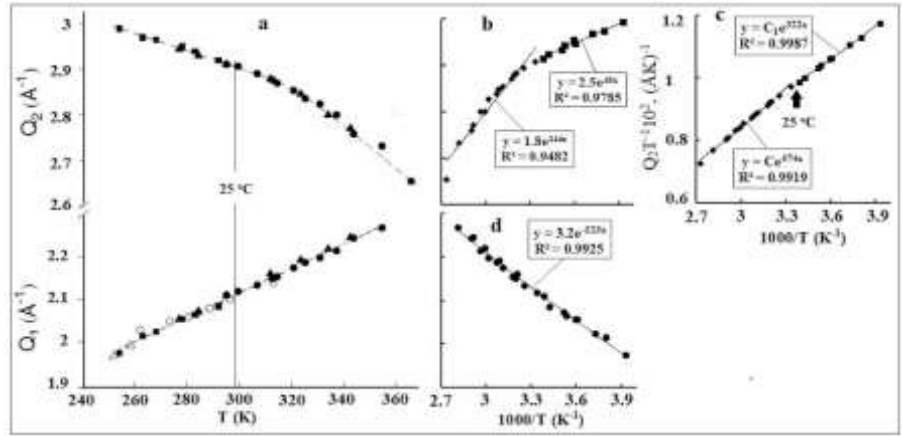

**Fig. 13.** (a) Temperature dependence of the first $Q_1$(Å$^{-1}$) and second $Q_2$(Å$^{-1}$) peak maximum positions in the measured x-ray water structure factor; (b, d) their $F_A$-approximations; (c) $F_R$-approximation. Initial data and adapted figure (a) from [15, 19, 21].

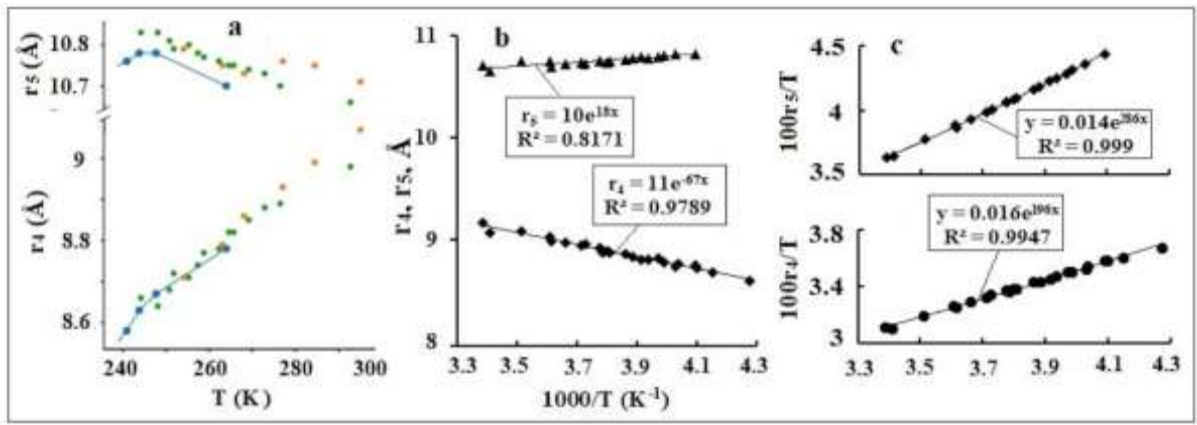

**Fig. 14.** (a) Peak location of the 4th ($r_4$~9Å) and of the 5th peak ($r_5$~11Å) of $g_{OO}(r)$. Their $F_A$-approximations (b) and $F_R$-approximations (c). Initial data and adapted figure (a) from [35].

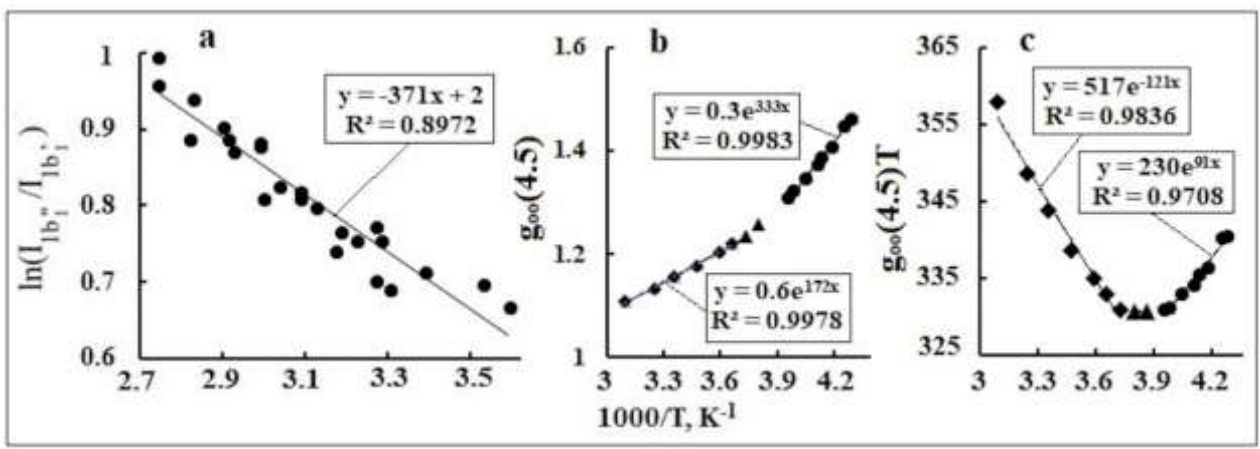

**Fig. 15.** (a) The temperature dependence of the logarithm of intensity ratio of bands 1b″ (distorted) and 1b′ (tetrahedral) of peaks in the lone-pair region of O in x-ray emission spectra of liquid $D_2O$ and its $F_A$-approximation. Initial data from [12]. (b) Dependence of the hight of peak $g_{oo}(4.5\ Å)$ on 1/T in x-ray scattering $H_2O$ and its $F_A$-approximation; (c) its $F_R$-approximation. Initial data from [21].

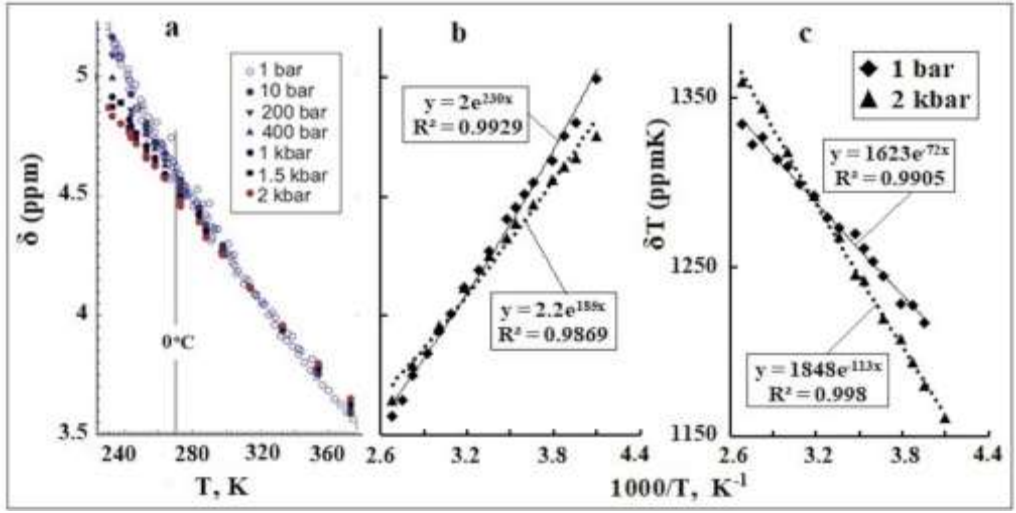

**Fig. 16.** (a) Temperature dependences of chemical shift (δ) for $H_2O$ at different pressures; (b, c) their $F_A$- and $F_R$-approximations. Initial data and adapted figure from [44];

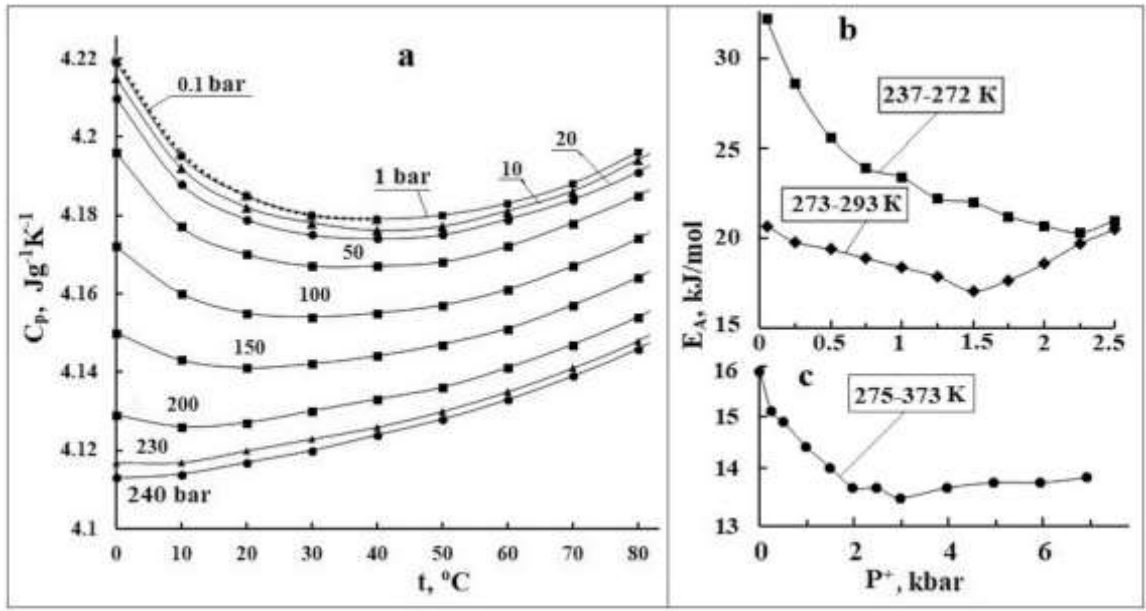

**Fig. 17.** (a) Temperature dependence of specific adiabatic heat capacity ($C_p$), for external pressure $P^+$ from 0.1 bar to 240 bars. (b) Dependence of activation energy ($E_A$) for spin-lattice relaxation ($T_1$) on pressure, in two ranges of T. (c) Dependence of $E_A$ for shearing viscosity (η) on pressure. Initial data (a), (b) and (c) are from [40], [41] and [9], respectively.

### 4. Discussion

By comparing the obtained values of $E_A$ and $E_R$ with $E_T$ (see Figure 5, Tables 1 and 2) and with the average value $E_H = 19$ kJ/mol [9, 25] it has come possible to divide the entire row of characteristics of liquid water in the following three, formally defined, groups: D, η, $τ_D$, $T_1$, P, f($ν_{oH}$) (1st group); γ, V, ε, ϰ, δ (2nd group); ρ, v, $C_p$, n (3rd group). The trends of energy activation variation in each group were analyzed in order to define specific features of molecular physics responsible for abnormal behavior of parameters of water.

Extreme points and activation energy of the temperature dependences
of the water spectral and structural characteristics.

| Spectral and structural parameters | | β | $t_E$ | Δt | $E_R$ | $E_A$ ($E_R + E_T$) | Fig. N; [Ref] |
|---|---|---|---|---|---|---|---|
| | | | | °C | kJ/mol | | |
| $\nu_{OH}$ (cm$^{-1}$) | 3260 | 0 | - | -2 – 47 | -18 | | 12 [26] |
| | 3620 | | | | 20 | | |
| Ln(1b″/1b′) | | | | 4 – 90 | 3.0 | | 15 [12] |
| $r_{oo}$ (Å) | 2.8 | 1 | 25 | -19 - 93 | 2.5 | -0.1 | 11 |
| | 4.5 | | | 0 – 25 | 2.3 | -0.06 | |
| | | | | 29 – 91 | 2.2 | -0.48 | |
| | ~9 | | - | -29 – 22 | 1.6 | -0.6 | 14 |
| | ~11 | | | | 2.4 | 0.15 | |
| Height $g_{oo}$ (4.5 Å) | | -1 | -12 | -40 – -15 | 0.8 | 2.7 | 15 |
| | | | | 0 – 50 | -1.0 | 1.4 | |
| $Q_1$ (Å$^{-1}$) | | 0 | - | -20 – 93 | -1.0 | | 13 [15], [19], [21] |
| $Q_2$ (Å$^{-1}$) | | 1 | 25 | -20 – 22 | 2.7±0.3 | 0.4±0.1 | |
| | | | | 26 – 93 | 3.9 | 1.2 | |

### 4.1. Characteristics D, η, $\tau_D$, $T_1$, P and $\nu_{OH}$

For the 1$^{st}$ group of characteristics, the following relationship is valid: $E_A \sim E_R \sim E_H \gg E_T$, in the range of 0 °C to 25 °C. Value of $E_A$ decreases by ~15% to 30%, in interval T > 25 °C, while in interval T < 0 °C it grows by ~20% to 50%. TDs do not have extremal points, while kinks of curves appear only in $F_A$-approximations, on the borders of these intervals. Dependence of pressure (P) on T at constant volume, is limited by endothermic reaction whose thermal effect is close to the breakage energy of HB that equals to $E_H$. This reaction is, actually, comparable to reactions of ice melting and formation of free molecules. In fact, TD of a portion of free molecules has $E_A = -7.6$ kJ/mol [5]. On the other hand, TDs for characteristics of η, $\tau_D$ and $T_1$ are limited by exothermal reactions with $E_A$ that are close to $E_H$, as well. Values of $E_A$ for TD of the average number of HB per one molecule are equal to 17.2 kJ/mol and 11.4 kJ/mol, in T-intervals 0° C to 25 °C and 30 °C to 90 °C, respectively [5], while rough estimate of $E_A$ for TD of number HBs per one molecule of water yields value 1.3 kJ/mol (see Fig. 8a). Thus, reactions responsible for TDs of characteristics η, $\tau_D$ and $T_1$ are, in principle, similar to reactions of HB formation in the process of water freezing.

The dynamics of HBs has its specific features in supercooled water state at T < 0 °C and may change considerably during $W_6$ clusters breaking, near 25 °C. These changes make their effect on the value of $E_H$, and they explain the reduction of certainty of $F_A$-approximations along with the growth of $E_A$, in intervals $T < 0$ °C, as well as the reduction of $E_A$, in $T > 25$ °C intervals (see Fig. 5, 11 to 14 and 16).

**Table 2**

Extreme points, parameter β and activation energy of Arrhenius approximations of the temperature dependences of the water characteristics.

| Fig. N, [Ref] | Water characteristics | | β | $t_E$ | Δt | $E_T$ | $E_R$ | $E_A$ ($E_R + E_T$) |
|---|---|---|---|---|---|---|---|---|
| | | | | | °C | | kJ/mol | |
| [5] | D (cm² s⁻¹) | | 1 | 0; 25 | -30 – 0 | -2.2 | -27.8 | -30 |
| | | | | | 0 – 25 | -2.7 | -17.5±0.4 | -20.2±0.3 |
| | | | | | 26-100 | -2.8 | -14.0±0.5 | -16.8±0.4 |
| [5] | η (cP) | | 0 | 0; 25 | -10 – 0 | -2.6 | 22.4 | |
| | | | | | 0 – 25 | | 19.0±0.3 | |
| | | | | | 25 – 100 | | 14.0 | |
| 4 | $τ_D$ (s) | | | 0 | -22 – 0 | | 25.5 | |
| | | | | | 2 – 100 | | 18 | |
| 6 | P (bar) | | 0 | - | 9 - 97 | | -18.7 | |
| 9 | ε | | -1 | 25; 46; 75 | 0 – 25 | 2.4 | 0.6 | 3.0 |
| | | | | | 30 – 45 | 2.6 | 1.1 | 3.7 |
| | | | | | 50 – 75 | 2.8 | 1.5 | 4.3 |
| | | | | | 76 – 95 | 3.2 | 2.1 | 5.3 |
| 8 | ϰ (Wm⁻¹ K⁻¹) | | 1/3 | ~72 | 2– 67 | -0.8 | -1.9 | -2.8 |
| | | | | | 72 – 97 | -1 | -0.6 | -1.6 |
| [5] | ρ (kg m⁻³) | | -1 | 0; 4 | -30 – 0 | 2.1 | -2.4 | -0.3 |
| | | | | | 4 – 100 | 2.7 | -2.3 | 0.4 |
| 6 | v (cm³ mol⁻¹) | | 1 | 4 | 4 – 100 | -2.65 | 2.28 | -0.37 |
| 3 | $C_p$ (J g⁻¹ K⁻¹) | Ice | 0 | - | -93 – -3 | - | -1.8 | |
| [7, 40] | | Water | 1 | 0; 36 | -27 – 0 | -2.1 | 4.5 | 2.4 |
| | | | | | 0 – 36 | -2.4 | 2.57 | 0.17 |
| | | | | | 37 – 100 | -2.81 | 2.68 | -0.13 |
| [5] | V (m s⁻¹) | | -0.5 | 25; 46; 75 | 0 – 25 | 1.1 | -2.9 | -1.8 |
| | | | | | 26 – 46 | 1.3 | -2.3 | -1.0 |
| | | | | | 47 – 75 | 1.38 | -1.73 | -0.35 |
| | | | | | 76-100 | 1.49 | -1.14 | 0.35 |
| 10 [5] | γ (bar⁻¹) | | 2 | 0; 25; 46; 75 | -30 – 0 | -4.4 | 10 | 5.6 |
| | | | | | 0 – 25 | -4.7 | 7.9 | 3.2 |
| | | | | | 26 – 46 | -5.1 | 6 | 0.9 |
| | | | | | 47 – 75 | -5.5 | 4.5 | -1.0 |
| | | | | | 76 – 100 | -6.0 | 3.1 | -2.9 |
| 16 | δ (ppm) | 1 bar | -1 | - | -29 –100 | -2.5 | -0,6 | 1.9 |
| | | 2 kbar | | | | -2.4 | -0,9 | 1.5 |
| 10 | n | | -0.5 | | 0 – 60 | 1.24 | -1.18 | 0.06 |

Opposite signs of $E_A$ for reactions responsible for TDs $f(\nu_{oH})$ and the fact that their values are close to that of $E_H$ make evident dependence of $\nu_{oH}$ on HB configurations with adjacent molecules. It also proves the assumption that reactions of HBs reconfiguration in SMS may be of both endothermic and exothermic type. Values $E_A$ at $T_1$, for sea water in normal conditions do not, practically vary (see Fig. 5). The matter is that in sea water ($C_{NaCl} \sim 0.6$ mol/l) with homogeneous distribution of Na and Cl ions, it follows from (2) that L ~ 11Å. It means that effects of ion hydration may, in principle, affect the dynamics of only 5$^{th}$ coordination sphere (see Table 1). Therefore, TDs for properties of physiological fluids ($C_{NaCl} \sim 0.16$ mol/l) in temperature interval ~40 °C >T>0 °C will be close to those of pure water. The distance between ions is L~6 Å, in electrolytes ($C_{NaCl} \sim 4$ mol/l), and hydration effects are comparable with the effect of high values of $P^+$ [38, 39]. This explains the observed variations of functions for matching distributions of the first and second coordination spheres [38], as well as the reduction of certainty of $F_A$-approximations for TDs of electrolyte characteristics.

External surplus pressure ($P^+$) belongs to mechanical factors producing stress, shearing and deformation, in water. The threshold for molecular motion excitation decreases as a result of $P^+$ effect, and the effect of friction grows simultaneously. As a consequence, values of $T_E$ and $E_A$ decrease the more the higher is $P^+$ and the lower is $E_A$ (see Fig. 17). High values of $E_A$ for characteristics within the 1$^{st}$ group compensate the effect of $P^+$ in temperature range $T > 0$ °C (see Fig. 4b), and the difference in the effect of $P^+$ on TDs of these properties becomes apparent only in minimum points of dependences $E_A$ on $P^+$, that is 1.5 kbar, for $T_1$, and 3 kbar, for η (see Fig. 17). One may assume that in the vicinity of $E_A$ the minimum threshold and inhibiting effects of $P^+$ become comparable with each other, and prevalence of inhibiting effect leads to the growth of $E_A$, in the range of higher $P^+$ values. In supercooled water, correlated growth of inhibiting effect and $E_A$ values can be observed. It is known [48] that the influence of high $P^+$ on translational diffusion (D and η) is much stronger than that on rotational one ($\tau_D$ and $T_1$). That accounts for the shift of $E_A$ minimum near $T_1$ from 1.5 kbar, in ordinary water, to 2.25 kbar, in supercooled one (see Fig. 17b).

### 4.2. Characteristics γ, V, ε, ϰ, δ and n

Specific feature of molecular physics responsible for TDs of characteristics γ, V, ε and δ is isotropic and equilibrium molecular dynamics limited by motions directed along an external force vector. In case of γ and V, it is mechanical factor similar to surplus pressure ($P^+$), while in case of ε and δ these are molecule polarization effects occurring due to the local electric and magnetic fields. Six degrees of freedom in translational and rotational self-diffusion are equivalent to each other [9], and six factoros correspond to them with activation energies $E_A$ whose values for D, η and $\tau_D$ are close to each other (see Table 2, [9]). Assuming that motions and intermolecular interactions collinear with the vector of external force are mainly responsible for TDs of γ, V and ε we have to consider just one of these six $E_A$ factors. In fact, values $|E_A/6|$ for D, η and $\tau_D$ correlate with those of $|E_A|$ for γ, V and ε, in temperature interval 0 °C to 25 °C which is far from the vicinity of $T_E$ for characteristics γ and V (see Table 2). Relationship $E_A^\gamma \sim 2E_A^V$ is in line with periodicity of acoustic wave effect that produces reversible compression of water structure. It follows from relationship $E_A^D \sim 6E_A^\varkappa$ (see Table 2) that a vectored flux of IR quanta produces anisotropic effect on the dynamics of water, quite similarly to acoustic waves. Unlike kinetic characteristics $\tau_D$ and $T_1$, value of δ depends on the magnetic field screening degree of a central molecule due to the effect of local magnetic fields produced by electrons inhabiting the closest shells. This dependence is, evidently, sensitive to the direction of constant external magnetic field. It depends on T in a similar manner to TD for ε. In this case, relationship $2E_A^\delta \sim E_A^\varepsilon$ can be attributed to the difference between orientation effects of orbital angular momenta of electrons and molecular dipole momenta. Thus, we can state that, in temperature range 0 °C to 25 °C, specific features of TDs for characteristics of the 1st and 2nd groups have an identical nature that can be, obviously, attributed to the transformations of $W_6$ within the lattice-type structure and to the transition between HDL and LDL phases, in liquid water [13, 44]. Participation of $W_6$ in this transition is confirmed by the presence of kink on TD curves for characteristics D, η, $T_1$, γ, V and ε at ~25 °C [4, 5] (see Table 2 and Figs. 5 and 9). Such kink is also present on TD plots for the first coordination sphere parameter $Q_2$(Å$^{-1}$) (see Fig. 13) and for the radius of the second sphere ($r_2$ = 4.5 Å) (see Figs. 11 and 12). TDs of these microstructure parameters are attributed to the transition between LDL and HDL phases of water [5, 8, 11, 13, 22].

Absolute $E_A$ value, for ε in temperature range of 0 °C to 25 °C, equals to $E_A$ = –3 kJ/mol, for ln(1b″/1b′) (see Table 2). According to the data presented in [10, 12, 17], $E_A'' \gg E_A'$. That is why value –3 kJ/mol can be associated with endothermic molecule rotation reactions in the field of delocalized HBs. It is assumed, in this case, that tetrahedral configuration of LDL water transforms into distorted configuration of HDL water [17, 22]. Motions of such type may occur in match with exothermic reactions of dipoles reorientation followed by formation of domains which will contribute to polarization effects on TD for ε. It is clear from Figs. 9 and 10 that reduction of ε and n values differ ~100 times, for equal intervals of T. In case of ε, the scale of effects is defined by the size of either domains or the fifth coordination sphere (~20 Å [9, 16]). Evidently, averaged effect of polarization in the scale comparable with wavelength of light ~589 nm is hundreds of times lower. The same order of magnitude has the relationship between values $E_A$ for $\varepsilon$ and n (see Table 2).

### 4.3. Characteristics $\rho$, v and $C_P$

In points $T_E$ of TDs for characteristic $\rho$, v, $C_P$, $\gamma$ and V, absolute values of $E_T$ and $E_R$ are identical and, therefore, $E_A = 0$ (see Table 2). In such conditions, isoenergetic second order phase transitions take place in which certain SMS configurations change in a correlated manner, in the entire volume of water [8, 24]. Points $T_E$ for these characteristics are critical ones in which the amplitude of long-wave fluctuations of order parameter grows to an extent that they become independent on micro-parameters controlling short-range intermolecular interactions [30]. In this situation, fluctuations of order parameter initiate transitions in the macro-system built by energetically coherent states, and a new either stable or metastable phase occurs, in SMS. The borders of the critical area are defined by relationship $(T-T_E)/T_E \approx 10^{-2}$ that is valid for extrema points $T_E$ of TDs for $\rho$, $C_P$, $\gamma$ and V. Rebuilding of water structure takes place in temperature interval $|T-T_E| \sim 4$ K where value $E_A$ remains close to zero (see Fig. 3c).

Taking into consideration high sensitivity of O-H vibration frequency ($v_{oH}$) to energy variations of the closest HBs and forces of electric field induced by adjacent molecules [23, 26, 51-53] one may assume that $v_{oH}$ is the major parameter of order. The limits of its effect can be estimated with the use of formula deduced for effective radius of exchange forces having electromagnetic origin [54, 55]:

$$L = \hbar C/E.$$

If we chose value $E_A$ in a point close to $T_E$ that for $C_P$, as an example, equals to 15 J/mol (see Fig. 3c) we obtain L~1 mm.

Molecular mechanism of transitions in points $T_E$ is, in fact, is similar to the physics of homogeneous crystallization of water at 0 °C. The key aspect of water crystallization physics at 0 °C and normal atmospheric pressure is exothermic chain reaction. As a result of this reaction, $W_6$ in active centers get integrated into lattice-type metastable structures (SMS*) that precede formation of ice Ih structuree. In paper [49] SMS* was called as metastable phase 'ice 0'. It was found out that each layer in such structure is a mirror image of the previous one. The major specific feature of active center is inhibiting reverse reactions leading to SMS* dissociation. It is achieved by transferring the thermal effect of crystallization reaction (Q = 6 kJ/mol) outside the borders of active centers. This energy value correlates with that of one IR quantum having $\omega$=500 cm$^{-1}$ or two quanta with $\omega$=250 cm$^{-1}$ (see Fig. 1). Such effective transfer of thermal energy Q takes place in water micro-drops, in experiment described in [15, 50] and in the course of snow formation, in the atmosphere, as well as in case of impurities (e.g. NaCl, $CO_2$) that reemit IR quanta (250 cm$^{-1}$ or 500 cm$^{-1}$), in SMS* spectral window.

By analogy with phase transition HDL→ LDL [10, 12], molecular physics of SMS* formation can be described as follows. In each coupled pair of $W_6$, reflection symmetric rotations with $E_A \sim$ –3 kJ/mol take place, in a correlated manner, along with the conjoint exothermic reactions leading to formation of HBs ($E_A \sim$ 6 kJ/mol) in which adjacent $W_6$ pair is involved. While heating from 0 °C to 4 °C, the reverse endothermic reaction LDL→ HDL develops that is accompanied by $W_6$ cells transformation, in SMS* layers. In this case, cavities occur that enclose free molecules, and the volumetric density and specific volume of water attain their maximum and minimum values, respectively. In the course of heating from 4 °C to ~25 °C transformation of SMS* in lattice-like HB structures in which $W_6$ 'monomers' prevail is activated. While heating over 25 °C $W_6$ transforms and dissociates into dimers and free molecules.

## 5. Conclusion

At normal pressure and temperatures in the range from 0 °C to ~100 °C, ambient water is represented by a solution of supramolecular structure (SMS). The lifetime, composition and degree of order of SMSs depend on temperature. Physics of liquid water combines the equilibrium thermodynamic self-diffusion and viscosity with phase transitions in SMS. Restructuring of the hydrogen bonds lattice and clusters within SMS is responsible for extrema of TDs for volumetric density and heat capacity, at constant pressure. In the vicinity of 25 °C, TDs for viscosity, self-diffusion, compressibility and certain microstructure parameters feature kinks of curve. Variation of molecular dynamics in this point is, evidently, defined by dissociation of the majority of ice-like hexagonal clusters within SMS. Thermal energy in TDs external points equals to the activation energy of motions responsible for phase transitions between SMS configurations of water. Extrema of TDs for compressibility and associated acoustical velocity are mainly determined by the effect of an external anisotropic factor on the equilibrium dynamics of water. A similar effect on the activation energy of water structure rebuilding produces external surplus pressure. Activation energies obtained in this work can be used as limiting conditions, in simulations of specific molecular dynamics of SMS water.

URL: http://www.sci.rostelecom67.ru/user/sgma/MMORPH/N-28-html/kholmanskiy-2/kholmanskiy-2.htm.